\documentclass[12pt]{iopart}

\usepackage{bm}
\usepackage{amssymb}
\usepackage[usenames,dvipsnames]{color}
\usepackage{graphicx}
\usepackage{hyperref}
\usepackage{soul}
\usepackage{float}
\usepackage{xspace}
\usepackage[group-separator={,}]{siunitx}
\usepackage{booktabs,tabularx,colortbl,multirow}
\usepackage{ifthen}
\usepackage{lineno}
\usepackage{xcolor}

\newcommand{\uiif}{Unpaired Image-to-Image\xspace}

\newcommand{\fake}{$A$\xspace}
\newcommand{\real}{$B$\xspace}
\newcommand{\faketoreal}{\fake~$\to$~\real\xspace}
\newcommand{\realtofake}{\real~$\to$~\fake\xspace}

\newcommand{\tpc}{LArTPC\xspace}
\newcommand{\dst}{SLATS\xspace}

\newcommand{\acl}{{ACL-GAN}\xspace}
\newcommand{\cyc}{{CycleGAN}\xspace}
\newcommand{\uga}{{U-GAT-IT}\xspace}
\newcommand{\uvc}{{UVCGAN}\xspace}

\newcommand{\genab}{\mathcal{G}_{A \to B}}
\newcommand{\genba}{\mathcal{G}_{B \to A}}
\newcommand{\disa}{\mathcal{D}_{A}}
\newcommand{\disb}{\mathcal{D}_{B}}

\newcommand{\norm}[1]{\left\Vert#1\right\Vert}
\newcommand{\paren}[1]{\left(#1\right)}
\newif\ifhighlighted
\highlightedfalse

\ifhighlighted
\definecolor{hlcolor}{rgb}{0.1, 0.4, 1.}
\else
\definecolor{hlcolor}{rgb}{0., 0., 0.}
\fi

\DeclareRobustCommand{\myhl}[1]{{\color{hlcolor}#1}}
\newif\ifblinded
\blindedfalse
\sethlcolor{lightgray}
\newcommand{\blinded}{\hl{[\texttt{Blinded for peer review}]}\xspace}
\def\paperTitle{Unpaired Image Translation to Mitigate Domain Shift in Liquid Argon Time Projection Chamber Detector Responses}
\def\paperTitleShort{UI2I translation to Mitigate Domain Shift in \tpc Detector Responses}

\def\authorList{
	\ifblinded
	\blinded
	\else
    Yi Huang, Dmitrii Torbunov, Brett Viren, Haiwang Yu, 
    Jin Huang, Meifeng Lin, Yihui Ren\footnote{corresponding author. Email:\url{yren@bnl.gov}.}\\
    Brookhaven National Laboratory, Upton, NY, USA\\
    {
        \tt\footnotesize 
        \{yhuang2,dtorbunov,bviren,hyu,jhuang,mlin,yren\}@bnl.gov
    }
	\fi
}
\begin{document}

\title[\paperTitleShort]{\paperTitle}
\author{\authorList}
\date{}

\begin{abstract}

Deep learning algorithms often are developed and trained on a training dataset and deployed on test datasets.  
Any systematic difference between the training and a test dataset may severely degrade the final algorithm performance on the test dataset---what is known as the \textit{domain shift problem}. 
This issue is prevalent in many scientific domains where algorithms are trained on simulated data but applied to real-world datasets.
Typically, the domain shift problem is solved through various domain adaptation methods.
However, these methods are often tailored for a specific downstream task, such as classification or semantic segmentation, and may not easily generalize to different tasks. 
This work explores the feasibility of using an alternative way to solve the domain shift problem that is not specific to any downstream algorithm.
The proposed approach relies on modern \uiif (UI2I) translation techniques, designed to find translations between different image domains in a fully unsupervised fashion.
In this study, the approach is applied to a domain shift problem commonly encountered in Liquid Argon Time Projection Chamber (\tpc) detector research when seeking a way to translate samples between two differently distributed \tpc detector datasets deterministically.
This translation allows for mapping real-world data into the simulated data domain where the downstream algorithms can be run with much less domain-shift-related performance degradation.
Conversely, using the translation from the simulated data to a real-world domain can increase the realism of the simulated dataset and reduce the magnitude of any systematic uncertainties.
To evaluate the quality of the translations, we use both pixel-wise metrics and a downstream task to measure the effectiveness of UI2I methods for mitigating the domain shift problem.
We adapted several popular UI2I translation algorithms to work on scientific data and demonstrated the viability of these techniques for solving the domain shift problem with \tpc detector data.
To facilitate further development of domain adaptation techniques for scientific datasets, the ``Simple Liquid-Argon Track Samples'' (\dst) dataset used in this study is also published.
\end{abstract}

\maketitle

\section{Introduction}
\label{sec:intro}

\begin{figure}[t]
	\centering
	\resizebox{\textwidth}{!}{\includegraphics{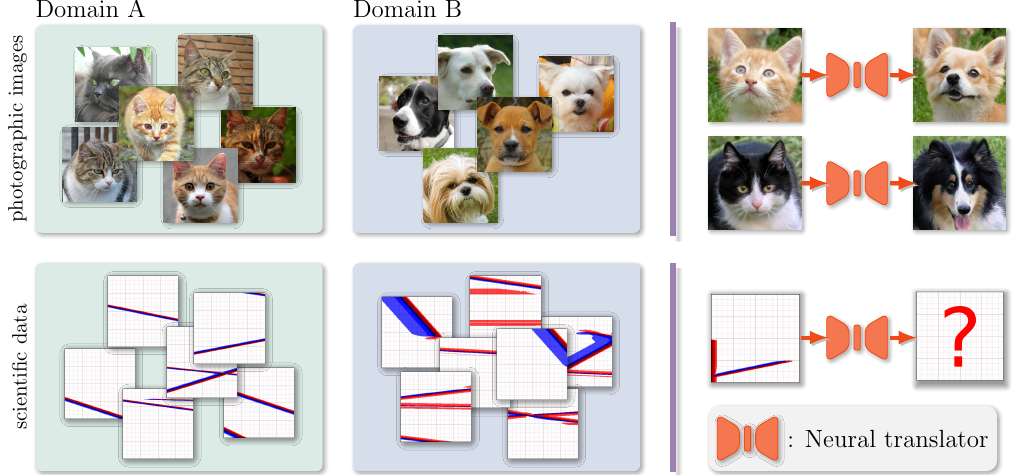}}    
	\caption{\textbf{Learning to translate without pairing.} An unpaired translation problem features two domains with samples that are not paired, e.g., cats and dogs. For an input image from the source domain, a neural translation algorithm needs to produce translations resembling samples in the target domain. In the meantime, the translations must retain certain consistency with their input. The first row demonstrates that a deep neural network model can be trained to translate cats into lifelike yet nonexistent dogs while maintaining features such as fur color patterns and facial orientations. Our work investigates if UI2I translation can be adapted to translate between two domains of \tpc images.}
	\label{fig:intro_could_images}
\end{figure}
\myhl{
Deep Learning (DL) methods are finding widespread and unprecedented applications in multiple areas of science and technology.
Constructing supervised DL models requires access to large volumes of properly \emph{labeled}, high-quality real-world data.
However, labeling real-world scientific data is often difficult, costly, or otherwise impossible~\cite{wang2018deep,dlforhealth,min2017deep,sapoval2022current}. 
To workaround the issue, many scientific domains resort to using simulation as a means of obtaining large quantities of labeled data.
Although this approach solves the lack of labeled data problem, it introduces another challenge. 
As there often exists systematic differences between the simulated and real-world data, 
a DL algorithm trained on a simulated dataset can exhibit degraded performance when it is applied to real-world data.
This issue is known as the \textit{domain shift problem}~\cite{DSinML,wang2018deep,takahashi2020review,fang2022source}.

In this work, we consider tackling the typical domain shift problem in Liquid Argon Time Projection Chamber (\tpc) detector research.
\tpc is a particle tracking and calorimetry detector technology~\cite{rubbia1977liquid,WILLIS1974221,Nygren:1974nfi} that forms the basis for detectors used by experiments such as MicroBooNE~\cite{acciarri2017design}, ProtoDUNE~\cite{abi2017single}, and the next-generation DUNE~\cite{abi2018dune}.
As with other detector technologies, obtaining human labels for real-world detector data is prohibitively costly.
Therefore, physicists rely on scientific detector simulations to generate labeled data and develop analysis algorithms.
% While experts regularly test manually designed analysis algorithms to minimize their vulnerability\textcolor{red}{How can one test an algorithm to minimize its vulnerability?} to the domain shift issue, this problem still poses a major challenge for data-driven deep learning algorithms. This has led to many debates, hindered the widespread use of deep learning methods in \tpc detector analysis, and driven efforts to find better solutions.
While the manually designed analysis algorithms are continuously tested to ensure they are minimally affected by the domain shift problem, domain shift remains a serious concern for DL algorithms. It has sparked numerous debates, significantly slowed adoption of DL methods to \tpc detector analysis, and driven the search for alternative solutions.

Typically, the domain shift problem is solved using various domain adaptation (DA) algorithms~\cite{csurka2017domain,fang2022source,wang2018deep,wilson2020survey}.
DA techniques are created to enable DL algorithms to perform effectively on novel domains distinct from those where they are trained.
However, \tpc data analysis workflows make it difficult to apply traditional DA techniques.
The primary obstacle is that state-of-the-art DA methods~\cite{guan2023iterative,hoyer2023mic,french2017self} are tightly coupled to a specific downstream task affected by the domain shift problem.
\tpc data analysis chains can employ dozens of different reconstruction algorithms possibly affected by the domain shift. 
This requires developing and testing dozens more DA methods, i.e., one for each downstream algorithm.
Moreover, new \tpc data analysis algorithms are constantly being developed, which requires designing even more new DA methods. 
Thus, it is not feasible to directly apply the traditional DA approaches to \tpc data analysis.

Here, we consider the viability of using \uiif (UI2I) translation methods to address the domain shift problem on \tpc data.
UI2I translation methods are developed for finding translations between different domains of images in a fully unsupervised way~\cite{liu2017unsupervised,zhu2017unpaired,torbunov2022uvcgan,aclgan,ugatit}.
For instance, the top row of \autoref{fig:intro_could_images} illustrates the operation of a UI2I translation algorithm for the cat-to-dog translation.
In the training phase, a UI2I translation algorithm receives random images from the two domains: cats and dogs. 
Notably, the UI2I translation algorithm is not given what exactly the correct translation of a particular cat should look like. 
Thus, the algorithm is called ``unpaired.''
Instead, the algorithm attempts to find some common ``content'' between the two domains and learns to perform a cat-to-dog translation while preserving the ``content.''
Once the algorithm is trained, it can transform an arbitrary image of a cat into an image of a dog
% that still resembles the input cat.
where the original cat and generated dog are related on some fundamental level (share the same ``content'').

The key question we try to answer is whether the UI2I translation methods are capable of learning the proper ``content'' and finding the ``correct'' translation between two domains of \tpc data:
domain $A$ representing simulation and domain $B$ denoting the real-world data.
This question is nontrivial as there is an infinite number of possible translations between the two domains mathematically, while only a small fraction of them is correct.
The UI2I translation literature frequently overlooks the question of whether or not the image content is preserved
during the translation or relies on subjective measures of the content~\cite{zhao2022egsde}.
Applying the UI2I translation methods to scientific data requires much stronger guarantees of the translation's correctness.
In this work, we investigate the ability of several UI2I translation models to learn the \tpc translations and compare their performance.

If the UI2I translation methods are capable of finding the correct translations between the data domains,
then the $B\to A$ translation can be used for DA purposes.
For instance, if an algorithm $\phi$ is trained on the $A$ domain and applied to an image $b$ from the $B$ domain, the domain shift problem would manifest.
However, if image $b$ is first translated toward the domain $A$ with the help of a UI2I translation algorithm, $\phi$ can be applied on the translated image, mitigating domain shift effects.

On the other hand, the correct $A \to B$ translation can be used to enhance the realism of the simulated data.
Performing such a translation on a simulated $A$ dataset will produce a more realistic $B'$ dataset, which has several potential applications:
\begin{itemize}
	\item It can be used to increase the fidelity of the simulation for the subsequent data analysis.
	\item \tpc analysis algorithms can be developed on the translated $B'$ data instead of the original simulated $A$ data. This has the potential to make the algorithms less affected by the domain shift effects.
	\item The $B'$ dataset is produced from $A$ by a UI2I translation algorithm.
		Therefore, for each $b'$ in $B'$, we know its source image $a$ in $A$.
		Comparing $b'$ to its source $a$ will allow for directly observing systematic differences between the simulation and real-world data on a sample-by-sample basis.
		Without such an $A \to B$ correspondence, scientists can only observe systematic differences by comparing averages over the entire dataset.
	\item Likewise, an $A \to B$ pairing can be used to estimate the sensitivity of various downstream algorithms $\phi$ to the systematic differences between the simulation and real-world data by computing $\phi(a) - \phi(b')$.
\end{itemize}

Unfortunately, it is difficult to evaluate UI2I translation methods on real data. 
For this to be possible, for each real detector data image, a matching simulation image should be generated with the same physics. Afterwards, the simulated image would be translated to see if the outcome matches the matching real image.
However, extracting the physics ground truth, such as particle momentum, from real-world \tpc data requires meticulous and time-consuming analysis from a large scientific collaboration.
Thus, we consider a \textbf{surrogate problem}, \emph{where both $A$ and $B$ domains are populated by the simulated data with controllable differences between the domains}.
Using the simulation will allow for making accurate judgments about the translation quality.

For this study, we created the Simple Liquid-Argon Track Samples (\dst) dataset featuring two domains: \fake and \real. 
The \fake domain is populated by a \tpc detector simulation with a simplified version of the detector response to the particle activity within it. 
The $B$ domain is populated by a \tpc detector simulation with a more realistic version of the detector response.
Incorrect simulation of the detector response is a known source of systematic errors in the \tpc detector analysis. 
Thus, the \dst dataset illustrates a common source of the domain shift problem encountered in \tpc detector research.
To facilitate the accurate evaluation of the translation's accuracy, the test portion of the \dst dataset has an explicit pairing between the \fake and \real domains. 
That is, for each test image in the \fake domain, we know exactly how its translation should appear in the \real domain and vice versa.

We evaluate the correctness of the resulting UI2I translations via three methods.
First, we use the explicit pairing of the test part of the \dst dataset to perform pixel-wise comparisons of the translated images to their ground truth.
Second, we rely on a downstream production-grade signal processing algorithm to extract the physical content of the images.
This algorithm is especially sensitive to the domain shift problem. 
%Lastly, we train several supervised learning algorithms to count the total number of electrons in different domains and verify that UI2I methods indeed mitigate the domain shift problem.  
Finally, we study whether the UI2I methods can improve the performance of a supervised DL algorithm affected by the domain shift on the \dst dataset.
% As such, it will help verify whether or not performing the UI2I translation mitigates the problem.

The remainder of this paper is organized as follows. In \autoref{sec:data}, we briefly describe how a \tpc detector works and the construction of the \dst dataset. 
In \autoref{sec:models}, we review a selection of UI2I translation algorithms suitable for \tpc data. 
In \autoref{sec:eval}, we evaluate the quality of translated images. Finally, in the discussion section, we summarize our findings and suggest future directions of research. 

\section*{Main Contributions}
\begin{itemize}
	\item We show the feasibility of using UI2I translation techniques to perform domain translation on \tpc data. 
		The four UI2I translation algorithms studied in this work manage to correctly capture the ``content'' of the data and preserve it during the domain translation.
	\item We demonstrate that UI2I translation techniques can be used to reliably mitigate the domain shift effects on \tpc data and can provide up to $80\%$ reduction in domain shift error of a downstream signal processing algorithm.
	\item Likewise, we show that UI2I translation methods can be used to improve the realism of the \tpc simulation, suggesting the viability of using UI2I translation methods as a post-processing step to obtain a more realistic simulation.
	\item We release a \dst dataset demonstrating the common source of the \tpc detector domain shift in a controllable manner.
		The dataset also displays unique features of scientific datasets not commonly shared by natural images, such as signal sparsity, lack of upper/lower limits on pixel values, and exact knowledge of the correct translations.
		This dataset is expected to help with the future development of additional scientifically sound domain translation algorithms.
	\item Finally, we compare the translation performance of four UI2I translation methods (\cyc~\cite{zhu2017unpaired}, \acl~\cite{aclgan}, \uga~\cite{ugatit}, \uvc~\cite{torbunov2022uvcgan}) on the \dst dataset. 
		Our results demonstrate that the \uvc algorithm significantly outperforms other methods across a wide range of metrics, and introduces the least amount of artifacts into the translated data.
		These results suggest that \uvc may serve as an effective basis for further development of UI2I translation methods in scientific data processing.
\end{itemize}
}

\section{The two-domain SLATS dataset}
\label{sec:data}

\begin{figure}[t]
	\centering
	\includegraphics[width=0.95\textwidth]{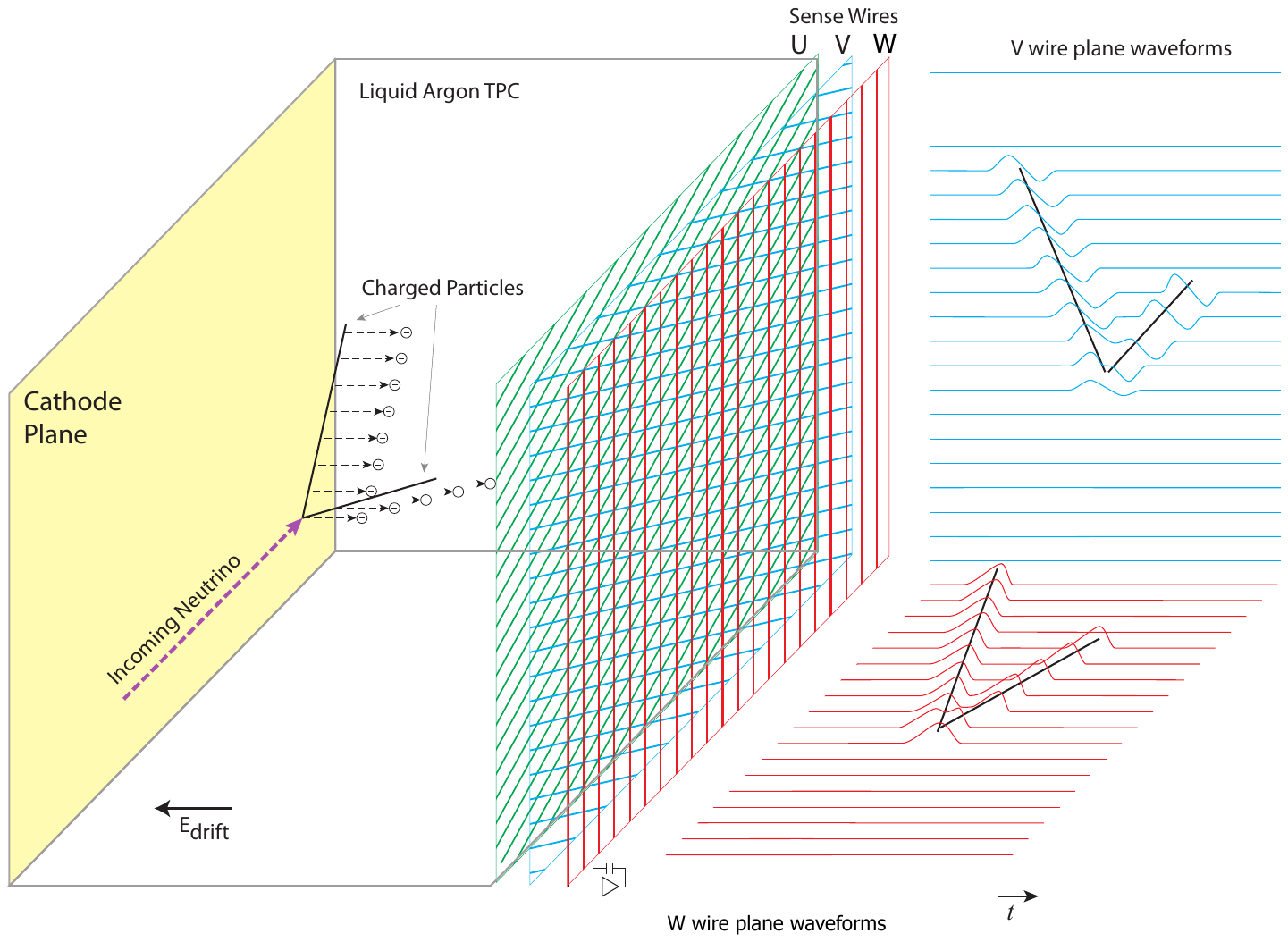}
	\caption{\textbf{Signal formation in a three-wire plane \tpc.} 
		An illustration from~\cite{MicroBooNE:2016pwy}. 
		\tpc detectors enclose a volume of liquid argon. 
		Energetic charged particles ionize electrons from nearby argon atoms as they pass through the volume. 
		An external electric field causes the electrons to drift toward the detector's readout. 
		The readout consists of three parallel planes of sense wires. Each wire plane generates one tomographic view of the tracks. 
		The 3D particle tracks can then be reconstructed by combining the three tomographic views.}
    	\label{fig:lartpc-sigform}
\end{figure}

\begin{figure}[t]
	\centering
	\resizebox{\textwidth}{!}{\includegraphics{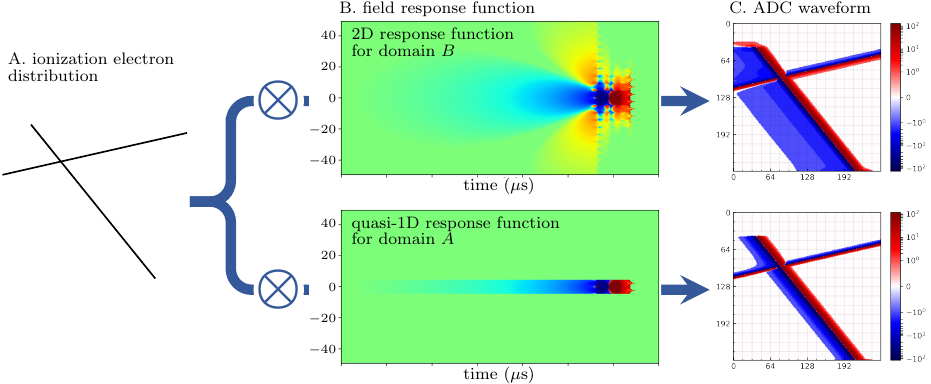}}
	\caption{\textbf{Response functions and ADC waveforms.} 
		The ionization electron distribution (Panel A) is convolved with two types of response functions to produce the \dst dataset's two domains. 
		The 2D response (Panel B top) is used to produce domain \real samples, while the quasi-1D response (Panel B bottom) is used to create domain \fake samples. 
		The quasi-1D data are constructed by masking the 2D response so all contributions from neighboring wires are removed. 
		Panel C shows examples of the ADC waveforms used as input to the translation algorithm.}
	\label{fig:data_simulation}
\end{figure}

\begin{figure}[t]
	\centering
	\resizebox{\textwidth}{!}{\includegraphics{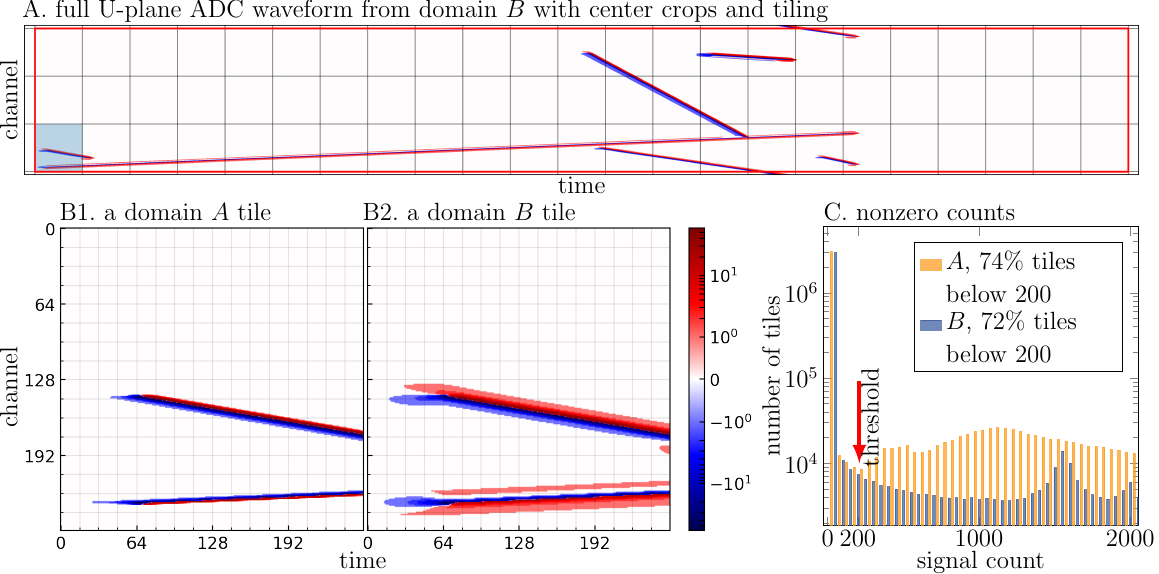}}
    	\caption{\textbf{Preprocessing for the \dst dataset.} 
		Panel A features an example of a full ADC waveform of the U plane from domain \real (generated with a 2D response). 
		The full image has dimension $(\text{channel}, \text{time}) = (800, 6000)$. 
		The portion bounded by the red box is the center crop of dimension $(768, 5888)$. 
		The center crop is divided into $3\times23$ tiles of size $(256, 256)$ and shown as the gray grid. 
		Panels B1 and B2 show a pair of tiles in the test dataset from the domain \fake (generated with a quasi-1D response) and the domain \real (generated with a 2D response), respectively. 
		The tile in B2 corresponds to the highlighted tile in Panel A. 
		The distribution of the number of nonzero pixels in the tiles is shown in Panel C. 
		Tiles with less than 200 nonzero pixels are discarded from the \dst dataset.}
	\label{fig:dataset_generation}
\end{figure}

Released with this study, the \dst dataset has two domains, each populated by a variant of a \tpc detector simulation used in the ProtoDUNE-SP experiment~\cite{pdsp-tdr,pdspres2020}.  
The two domains differ in precisely one feature---\textit{the response function}. This section discusses how the \dst dataset is generated and preprocessed.

\subsection{\tpc overview}
\label{subsec:data-tpcoverview}

\tpc detectors enclose a volume of liquid argon (\autoref{fig:lartpc-sigform}). 
Energetic charged particles, like those produced from the interaction between a neutrino and an argon nucleus, pass through the volume. 
As they move, these particles ionize electrons from nearby argon atoms. 
An external electric field causes these electrons to drift through the liquid argon toward the detector's readout side. 
The readout of the detector comprises three parallel planes (U, V, and Y) of parallel sense wires, oriented in complementary directions. 
Each wire plane generates a readout, called an \textit{ADC waveform}, that can be interpreted as one tomographic view of the particle's tracks. 
A tomographic view is a two-dimensional (2D) image with one dimension in space and the other in time. 
When the three tomographic views are combined, the three-dimensional (3D) tracks of the energized charged particles can be reconstructed.

In this work, the images used to construct the \dst dataset are the readout from one wire plane (the U plane). 
The pixel value of the images is the digitized measure (or ADC value) of the current induced by the ionized electrons. 
The measure is the result of a convolution between the electron distribution and a detector response~\cite{ramo1939}. 
The real detector response is a complex function of the electrostatic fields of all electrodes in the detector's readout. 
However, in simulation and signal processing, this response is approximated by a simplified model. 
We call such an approximation a \emph{response function}.

\subsection{Two simulated domains}
\label{subsec:data-simulated}

The two \dst dataset domains, \fake and \real, are generated by applying two different response functions. 
More precisely, for a set of simulated simple particle tracks, a low-fidelity quasi-one dimensional (1D) response function is applied to produce a domain \fake waveform and a high-fidelity 2D response function to produce a domain \real waveform. 
The 2D model is state-of-the-art in the \tpc community. 
The quasi-1D model is an artificially simplified model obtained by masking all contributions from neighboring wires in the 2D response (\autoref{fig:data_simulation}). 
Additional information on response functions and track generation can be found in 
Appendix A.1. 
% \ref*{supp:data-simulation}. 

The \dst dataset's design was motivated by three considerations. 
First, in the absence of real detector data, the contrast between the 2D model and the quasi-1D model is a reasonable proxy for the systematic difference between the real detector response and a simulated one. 
Second, using identical simulation conditions except for response functions allows for generating \emph{paired test images}. 
With the paired test images, we can evaluate a UI2I translation algorithm by directly comparing a translated image with its known target. 
Lastly, the restricted source of difference in the two domains facilitates understanding of the capability (and/or potential limitations) of unpaired neural translation. 
The experience gained via such a constructed scenario affords a proper foundation for applying UI2I translation between domains with complex sources of difference.

\subsection{Dataset preparation}
\label{subsec:data-preparation}

This study focuses on one of the three sense wire planes, namely the U plane. \autoref{fig:dataset_generation}A depicts a full U plane readout of dimension $(\text{channel}, \text{time}) = (800, 6000)$. 
Because a majority of existing neural translation algorithms take an input size of $(256, 256)$, we use a center $(768, 5888)$ crop (red box) in the U plane image, and then divide it into $3\times23$ non-overlapping tiles of size $(256, 256)$.

\autoref{fig:dataset_generation}B1 and B2 show a pair of tiles from domains \fake and \real. 
Two major differences appear between them. 
First, the domain \real track exhibits long-range induction effects in both the time and space (channel) dimensions, while the \fake track shows variation only in time. 
This leads to domain \real tracks being less compact than domain \fake. 
In particular, larger lobe structures can be observed at the end of domain \real tracks, while domain \fake tracks end more abruptly. 
This can also lead to features in a domain \real tile missing from the corresponding domain \fake tile, such as the small red lobe between the two tracks. 
Second, domain \real has a larger neighborhood where the electron distribution can lead to interference patterns as evidenced by the red lobe above the bottom track. 

Because of the sparseness of events in the generation of \dst, a majority of $(256, 256)$ tiles are fully or nearly empty. 
According to the distribution of the number of nonzero pixels in the tiles (\autoref{fig:dataset_generation}C), we choose a threshold of $200$ pixels (around the first local minimum for domain \fake) and reject tiles below the threshold. 
To keep the tiles paired for testing, we retain a pair in the test dataset if only both the \fake and \real tiles pass the threshold. More details about the preprocessing of \tpc simulation data for neural translator training can be found in  
Appendix A.2.
% \ref*{supp:data_preprocessing}. 

The \dst dataset can be downloaded from 
\ifblinded\blinded\else\url{https://zenodo.org/record/7809108}\fi.
The dataset contains both center crops and tiles. 
The dataset's test part is paired for pairwise translation quality evaluation.

\section{Deep generative models for unpaired image translation}
\label{sec:models}

As previously outlined, our goal is to apply modern UI2I translation methods to mitigate the domain shift problem. 
This section addresses the challenges in designing UI2I translation algorithms and describes a family of UI2I translation methods suitable for this task. 
The algorithms discussed herein are based on a Generative Neural Network (GAN) architecture (\autoref{subsec:model-gan}) and rely on a cycle-consistency constraint (\autoref{subsec:model-cycle}) to ensure preservation of the important features during the translation. 

\subsection{GAN for UI2I}
\label{subsec:model-gan}
The first successful models for UI2I translation were built on top of the GAN architecture~\cite{goodfellow2020generative}.
GAN models are able to learn the particular data distribution and synthesize new samples indistinguishable from the real data.
Their main component is a generator network $\mathcal{G}$ that produces realistic-looking data from random noise.
To train the generator network $\mathcal{G}$, GANs employ another neural network known as a discriminator $\mathcal{D}$.
In each GAN training iteration, the discriminator $\mathcal{D}$ network learns to differentiate samples produced by the generator from the real-world data.
Then, using the discriminator as a guide, the generator $\mathcal{G}$ network is trained to produce samples that are indistinguishable from the real-world data.
In other words, the generator and discriminator engage in a game, throughout which the generator progressively improves the quality of the generated data.

\subsection{Cycle-consistent GAN}
\label{subsec:model-cycle}

\begin{figure}[t]
	\centering
	\resizebox{.8\textwidth}{!}{\includegraphics{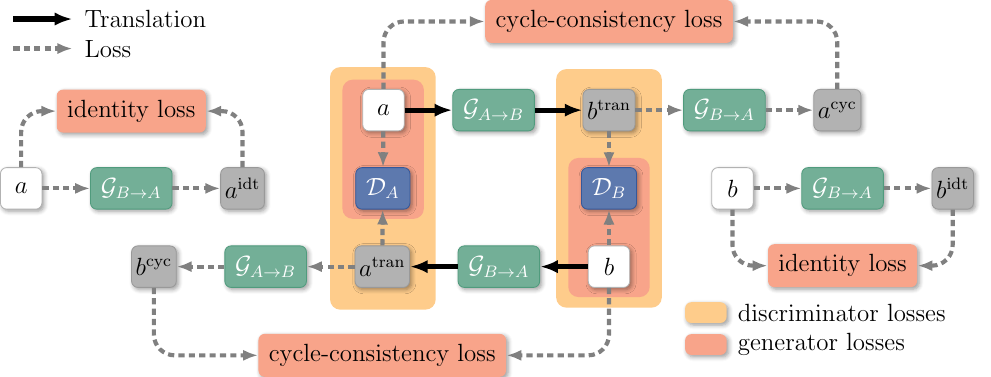}}
	\caption{\textbf{Summary of the \cyc~\cite{zhu2017unpaired} model.} 
		\cyc consists of two pairs of GANs, $(\genab, \disb)$ and $(\genba, \disa)$. 
		The discriminators, $\disa$ and $\disb$, distinguish translations from real images, while the generators (or translators), $\genab$ and $\genba$, produce realistic translations that are consistent with the source images.}
	\label{fig:model_CycleGAN}
\end{figure}

As noted, GANs can be used to learn data distributions and produce realistic-looking samples from random noise. 
This means, in principle, a GAN can be trained to generate real detector data from simulated data.
However, when a GAN generates a real detector data sample from a simulated one, it is not guaranteed to preserve any information from the input. 
The GAN can completely discard the simulated sample and produce a random and unrelated output that looks like real detector data.
This creates a challenge for our \tpc detector example as we need to generate not only realistic-looking \textit{real}
data samples, but also ensure the generated samples preserve information from the simulated ones.
The same discussion also applies to the translation in the opposite direction.

One approach to address this is provided by \cyc~\cite{zhu2017unpaired}, which employs \emph{two} GANs that work in the opposite directions as illustrated in \autoref{fig:model_CycleGAN}. 
By using a pair of GANs, a \cyc-like model creates translation loops, so information loss during translation can be properly measured.

Specifically, we denote the two domains by $A$ and $B$ and the corresponding GANs as $(\genab, \disb)$ and $(\genba, \disa)$, respectively. Consider a source image $a\in A$.
A \cyc-like model can translate this sample to look like those from domain $B$ by using its generator $\genab$.
To ensure the generator $\genab$ preserves information about the source image $a$, \cyc imposes a cycle-consistency constraint, requiring that a cyclically translated image $a^\text{cyc} \equiv \genba(\genab(a))$ matches the original image $a$.
In practice, this constraint is enforced by the \emph{cycle-consistency loss} $\norm{a - a^\text{cyc}}$ that encourages the generators $\genab$ and $\genba$ to preserve the information. 
A similar loss function is applied for the cyclic translation starting from $b \in B$.

In addition to the cycle-consistency loss, \emph{identity loss} may be used to encourage the generator to retain features from the source that are also present in the target domain.
For an image $a \in A$, the identity loss is defined as $\norm{a - a^\text{idt}}$, where $a^\text{idt}\equiv\genba(a)$.  A parallel formulation applies to domain $B$.

\subsection{\cyc-like UI2I translation models}
\label{subsec:model-cyclegan-like}

Here, we introduce three \cyc-like UI2I translation algorithms, \acl~\cite{aclgan}, \uga~\cite{ugatit}, and \uvc~\cite{torbunov2022uvcgan}, with special emphasis on the \uvc, or U-Net Vision-transformer Cycle-consistent GAN, because of its outstanding performance on the \dst dataset (see ~\autoref{sec:eval}). 

The motivation behind \acl is that the stringent pixel-wise cycle-consistency loss may be a hurdle for generators to produce drastic changes such as large shape changes or removing/adding large objects. To solve the problem, \acl replaces strong cycle-consistency loss with a weaker adversarial consistency that does not require the cyclically translated image to match the source exactly, merely to match the distribution of the source images.

The authors of \uga attack the problem of effective translation from another angle, keeping the cycle/identity-consistency loss functions in their original form as those in \cyc but renovating the generator and discriminator network structure. They use the class attention map to guide the generators and discriminators to focus on regions distinguishing between source and target domains. \uga has achieved outstanding performance in translation between selfie photos and anime characters, which is a tough image translation task. One downside of \uga is its model is bulky and slow, which may limit its application to \tpc research should throughput and computing resources become pressing considerations.

Based on this work, the \uvc model performs the best for translations between the two \dst dataset domains.
\uvc improves \cyc by renovating its generator network and the training procedure. 
The \uvc generator is a hybrid architecture of a U-Net backbone~\cite{ronneberger2015unet} with a Vision-Transformer (ViT) bottleneck~\cite{dosovitskiy2020image}. 
U-Net is known for its outstanding accuracy in modeling local or short-range patterns and its application in the segmentation of medical images. 
However, it may be less effective at capturing long-range dependencies.
Conversely, based on its impressive performance in image classification~\cite{anokhin2021image}, ViT excels in capturing long-range dependencies and semantic relationships within an image. 
Nevertheless, relying solely on ViT may be insufficient for addressing the complexity of an image translation task, a regression problem in nature, as it may struggle to model details.
Hence, the hybrid generator architecture of \uvc amalgamates the strengths of convolution-based networks and ViT, striking a balance between local and long-range pattern recognition.

\subsection{Alternative models for unpaired image translation}
\label{subsec:model-benchmarking}

Numerous models have been developed for UI2I translation, primarily on non-scientific image datasets. 
The models can be categorized based on two perspectives: the DL paradigm the algorithm is based upon and the way that consistency is enforced. 
For example, based on the paradigm, \cyc~\cite{zhu2017unpaired}, \acl~\cite{aclgan}, \uga~\cite{ugatit}, Council-GAN~\cite{council}, and \uvc~\cite{torbunov2022uvcgan} are GAN-based methods.
CUT~\cite{cut} adopts the contrastive learning paradigm. 
LETIT~\cite{letit} utilizes the energy transport on the latent feature space, while EGSDE~\cite{zhao2022egsde} and ILVR~\cite{choi2021ilvr} are diffusion-based models. 
In terms of consistency enforcement, \cyc, \acl, \uga, \uvc, and CUT impose explicit consistency constraints via loss functions, while the other methods do so implicitly. 

Another key feature of UI2I translation algorithms is the use of artificial randomness in the image generation process. 
Among the aforementioned models, \cyc, \uga, \uvc, and CUT are deterministic, while Council-GAN, EGSDE, and ILVR inject randomness into image generation. 
Although randomness helps boost diversity in natural image translation tasks, as there tends to be no single correct translation corresponding to an input, its application to \dst is unnecessary. Specifically, for this study of the idealized \dst dataset, the map between the two domains is \emph{one-to-one} in nature, which makes a deterministic model the more appropriate choice.

Given the limitation on time and computing resources, we focus on four models that enforce cycle consistency explicitly because models without explicit cycle consistency place virtually no constraints on the output and may generate images unrelated to the input.

\section{Evaluation}
\label{sec:eval}

As part of this work, the performance of the neural translation algorithms \cyc, \acl, \uga, and \uvc is evaluated on the paired test set of \dst.

First, we perform a direct pixel-wise comparison of the translated detector readouts (ADC waveform images) with their targets. 
This comparison will indicate the quality of translation on the raw detector readout level. Second, a signal processing algorithm (see
Appendix D, 
%\ref*{supp:sigproc-detail}, 
and \cite{microboonesigproc2018p1}) is applied to estimate physically meaningful counts of ionized electrons~(see \autoref{sec:data}) from the raw detector readouts. 
The signal processing algorithm is designed to perform accurately on domain \fake, and it exhibits domain-shift-related performance degradation when applied to the data from domain \real. 
Using the signal processing algorithm allows for estimating the degree to which the UI2I translation algorithms alleviate the domain-shift effects on physically meaningful quantities.

Of note, making \cyc, \acl, \uga, and \uvc work on the \dst dataset required several modifications. 
To explore their potential, we conducted a small-scale hyperparameter (HP) tuning on each of the algorithms. 
For simplicity, all results in this section are produced by the best-performing HP settings of each algorithm. 
The details regarding model modification, HP tuning, and training are available in 
Appendix B
%\ref*{supp:model-uvcgan-training} 
for \uvc and in 
Appendix C
%\ref{supp:model-other} 
for the other three \cyc-like models.

\begin{table}[t]
	% ================== parameters ==================
	\setlength\tabcolsep{5pt}
	\addtolength{\extrarowheight}{\belowrulesep}
	\aboverulesep=0pt
	\belowrulesep=0pt
	% ================== parameters ==================
    
	\centering
	\caption{\textbf{Translation performance comparison with $\ell_1$ and $\ell_2$ differences on the ADC waveform}. 
		The differences are produced with the best performer of each algorithm. Full results with all HP settings can be found in 
		Appendix Table E1.
		% \ref*{supp_tab:metrics}.
	} 
	\begin{tabular}{r|rr|rr}
		\toprule
		{} & \multicolumn{2}{r|}{\fake to \real} & \multicolumn{2}{r}{\real to \fake} \\
		Algorithm & $\ell_1$ & $\ell_2$ & $\ell_1$ & $\ell_2$ \\
		\midrule
		\cyc & $0.074$ & $0.180$ & $0.061$ & $0.159$ \\
		\acl & $0.083$ & $0.566$ & $0.039$ & $0.121$ \\
		\uga & $0.078$ & $1.187$ & $0.073$ & $1.161$ \\
		\textbf{\uvc} & $\mathbf{0.030}$ & $\mathbf{0.033}$ & $\mathbf{0.025}$ & $\mathbf{0.027}$ \\
		\bottomrule
	\end{tabular}
	\label{tab:metrics}
\end{table}

\begin{figure}[t]
	\centering
	\resizebox{\textwidth}{!}{\includegraphics{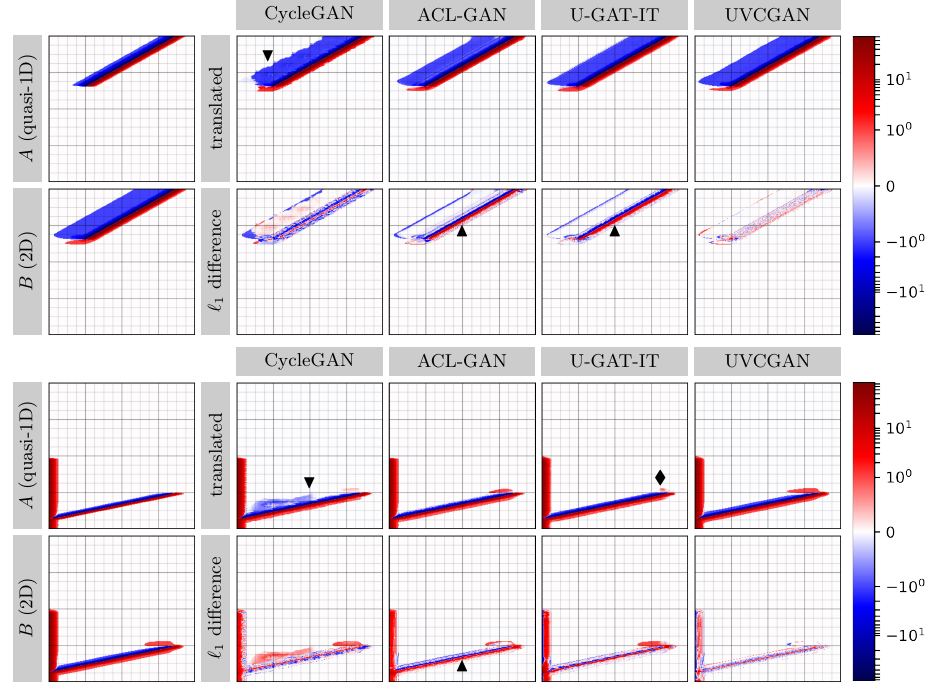}}
	\caption{\textbf{Examples for the \faketoreal translation.}
		Defects appearing in the translations are marked as:
		$\blacktriangledown$ for rugged track edge, 
		$\blacktriangle$ for big error in the core of the track where the signal is strongest, 
		and $\blacklozenge$ for missing the ``lobe'' structure near the track tip.}
	\label{fig:adc_grid_A2B}
\end{figure}

\begin{figure}[t]
	\centering
	\resizebox{\textwidth}{!}{\includegraphics{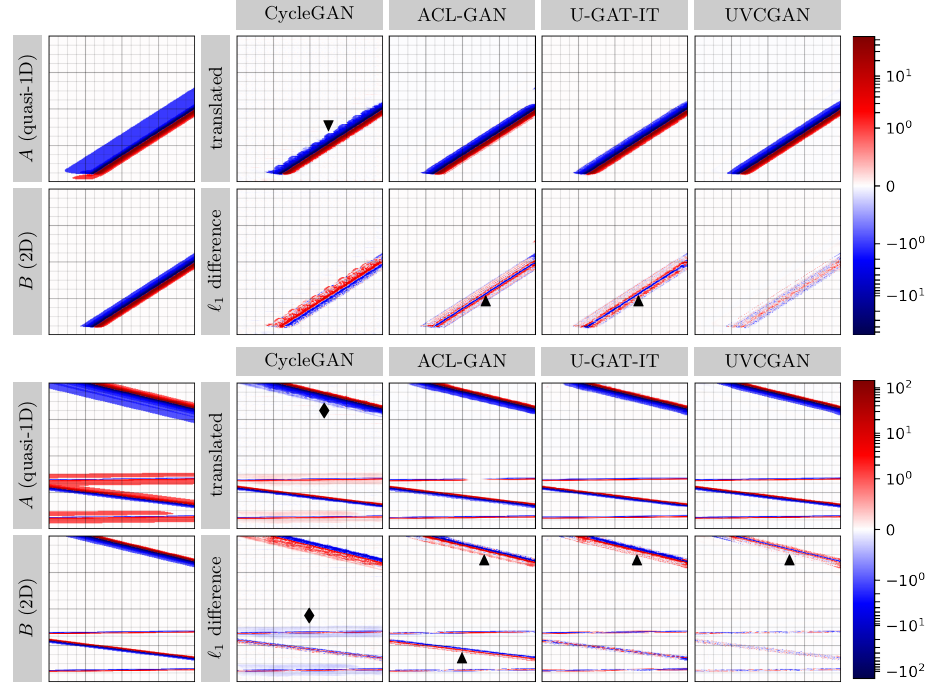}}
	\caption{\textbf{Examples for the \realtofake translation.}
		Defects appearing in the translations are marked as:
		$\blacktriangledown$ for rugged track edge,
		$\blacktriangle$ for big error in the core of the track where the signal is strongest, 
		and $\blacklozenge$ for incompletely reduced track edges.}
	\label{fig:adc_grid_B2A}
\end{figure}

\subsection{Translation quality evaluated on ADC waveforms}
\label{subsec:eval_metrics}

To quantitatively estimate the quality of the ADC waveform translations, we calculate $\ell_1$ (mean absolute error) and $\ell_2$ (mean squared error) between the translated and ground truth images. 
\autoref{tab:metrics} summarizes the best-performing results, while the complete results for all HP settings can be found in 
Appendix Table E1.
% \ref*{supp_tab:metrics}. 

Two samples from the \faketoreal translation in \autoref{fig:adc_grid_A2B} and another two from the \realtofake translation in \autoref{fig:adc_grid_B2A} are presented for a qualitative comparison, which shows all algorithms manage to reproduce the key features of the target domain in the translations to some extent. 
The features include more extended tracks and ``lobe'' structures at the track tips in the \faketoreal translation and compactified tracks and more abrupt track tips in the \realtofake translation (see~\autoref{subsec:data-preparation}). 
However, there are several noticeable defects in the translations, such as rugged track edges, large errors in the track center, missing ``lobes'' near track tips in \faketoreal translation, and incompletely reduced track edges in \realtofake translation. 
That said, all of the translation algorithms perform reasonably well in maintaining a strong consistency with the input images.

\begin{figure}[t]
	\centering
	\resizebox{\textwidth}{!}{\includegraphics{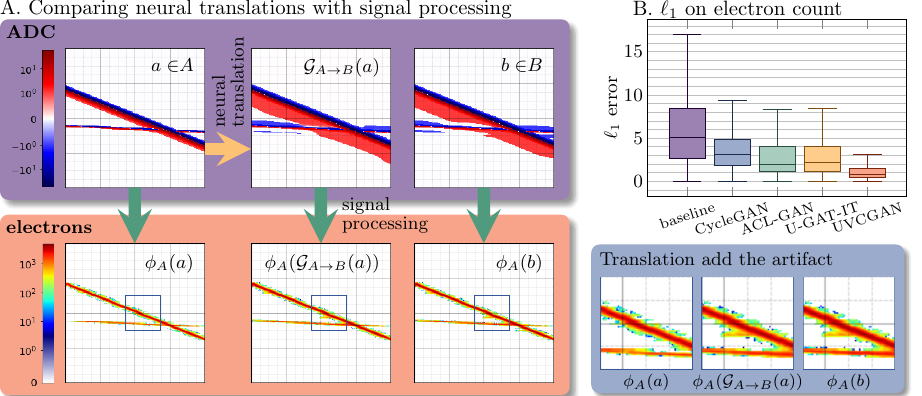}}
	\caption{\textbf{Signal processing study for \faketoreal translation.} 
		Panel A features a diagram of the signal processing study. 
		A tile $a$ from domain \fake is translated by the \uvc generator $\mathcal{G}_{A\rightarrow B}$ to $\mathcal{G}_{A\rightarrow B}(a)$, which resembles $a$'s counterpart $b$ from domain \real. 
		To reconstruct the electron count, signal processing $\phi_A$ is applied to $a$, $\mathcal{G}_{A\rightarrow B}(a)$, and $b$. 
		We zoom in on an area where $\phi_{A}(b)$ exhibits an artifact. 
		Because the artifact is absent from $\phi_{A}(a)$, we know it is a result of the mismatch between the response function and the signal processing procedure. 
		A similar artifact can be observed in the signal processed translation $\mathcal{G}_{A\rightarrow B}(a)$, which attests to the effectiveness of the translation.
		Panel B compares the $\ell_1$ errors on electron count. 
		Comparing the result with \autoref{tab:metrics} illustrates the translation quality in ADC values correlates strongly with post-signal processing performance.}
	\label{fig:sigproc_ab}
\end{figure}

\begin{figure}[t]
	\centering
	\resizebox{\textwidth}{!}{\includegraphics{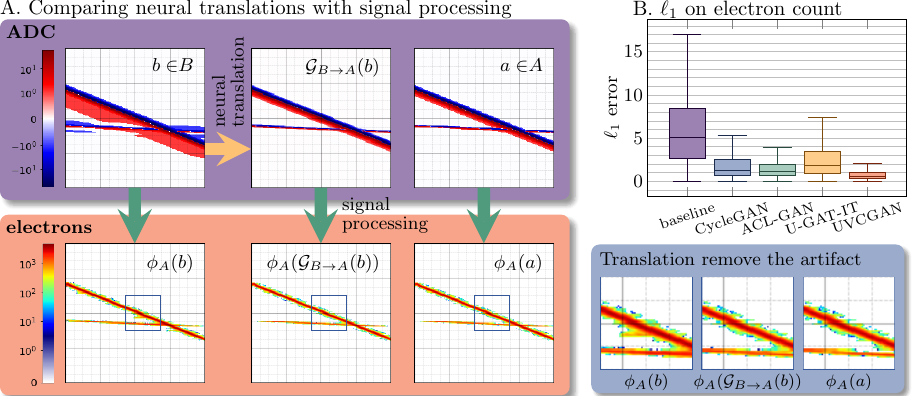}}
	\caption{\textbf{Signal processing study for \realtofake translation.} 
		Panel A depicts a diagram of the signal processing study. 
		A tile $b$ from domain \real is translated by the \uvc generator $\mathcal{G}_{B\rightarrow A}$ to an image $\mathcal{G}_{B\rightarrow A}(b)$, which resembles $b$'s counterpart $a$ from domain \fake. 
		Signal processing $\phi_{A}$ is applied to $b$, $\mathcal{G}_{B\rightarrow A}(b)$ and $a$, so the electron count can be reconstructed. 
		As in \autoref{fig:sigproc_ab}A, we zoom in on the same area where $\phi_{A}(b)$ exhibits an artifact. 
		This shows the artifact disappears in the signal processed translation $\mathcal{G}_{B\rightarrow A}(b)$, which attests to the effectiveness of the translation.
		Panel B compares the $\ell_1$ errors on electron count. 
		Comparing the result with \autoref{tab:metrics} demonstrates the translation quality in ADC values correlates strongly with post-signal processing performance.}
	\label{fig:sigproc_ba}
\end{figure}

\subsection{Translation quality evaluated on signal processing results}
\label{subsec:eval_sigproc}

The raw \tpc detector readouts are represented as ADC waveforms.
These waveforms are difficult to interpret and have no direct relation to the physical properties of the particles that created them.
Therefore, instead of building detector reconstruction pipelines directly on such waveforms, a signal processing algorithm is run first.
The signal processing algorithm is designed to infer the original, physically meaningful, distribution of ionized electrons that induced a particular waveform (cf.~\autoref{sec:data}).
The recovered distributions of ionized electrons serve as a basis for the downstream reconstruction algorithms.

The signal processing algorithm involves two main stages: deconvolution of an ADC readout and high-pass filtering.
The deconvolution operation is designed to act as an inverse of the simulated detector response function.
Therefore, it is affected by the domain shift, as the simulated detector response may differ from the real detector response.
Since signal processing is the first stage of the detector reconstruction pipelines, its domain shift error is then propagated to downstream algorithms.

The second stage of the signal processing algorithm is high-pass filtering.
It is required since the bipolar nature of the deconvolution operator tends to amplify low-frequency noise.
An adaptive high-pass filter, referred to as the \textit{signal region-of-interest} (\textit{ROI}) \textit{selection}, is subsequently applied to mitigate the impact of this amplification.
Further details of this algorithm can be found in 
Appendix D.
%~\ref*{supp:sigproc-detail}.

In this study, where domain \real is used as a proxy to real detector data, we naturally use the quasi-1D response function to design the signal process procedure and denote the signal processing as $\phi_A$. 
Because domain \fake has a matching simulation and signal processing, the electron counts $\phi_A(a)$ reconstructed from $a\in$\fake should match the ground truth electron counts (minus the random noise introduced in the signal processing procedure).
On the contrary, because domain \real is simulated and signal processed with different response functions, the electron counts $\phi_A(b)$ reconstructed from $b\in$\real will be less accurate than its counterpart $\phi_A(a)$. 
The difference between $\phi_A(a)$ and $\phi_A(b)$ is an indicator of the severity of the domain shift problem. 
Consequently, the reduction in the difference resulting from the translation can be viewed as the extent to which the domain shift problem is mitigated. 

To carry out a quantitative study, we randomly sample $1000$ pairs of test \dst tiles. 
We calculate the baseline $\ell_1$ error $\norm{\phi_{A}(a) - \phi_{A}(b)}_1$ and translation $\ell_1$ error $\norm{\phi_{A}\left(\mathcal{G}_{A\rightarrow B}(a)\right) - \phi_{A}(b)}_1$ and $\norm{\phi_{A}\left(\mathcal{G}_{B\rightarrow A}(b)\right) - \phi_{A}(a)}_1$, where $\mathcal{G}_{A\rightarrow B}$ and $\mathcal{G}_{B\rightarrow A}$ are neural translators. 
\autoref{fig:sigproc_ab}B and \autoref{fig:sigproc_ba}B reveal the statistics.

It is worth noting that because $\phi_A$ is designed based on the detector response function of domain \fake, only the comparison between $\norm{\phi_{A}\left(\mathcal{G}_{B\rightarrow A}(b)\right) - \phi_{A}(a)}_1$ and the baseline $\norm{\phi_{A}(a) - \phi_{A}(b)}_1$ can be used to infer the extent to which a translation can reduce the domain-shift effect. 
However, $\norm{\phi_{A}\left(\mathcal{G}_{A\rightarrow B}(a)\right) - \phi_{A}(b)}_1$ also is a valid indicator of the translation efficacy as it measures the translation's sensitivity in capturing the inaccuracy resulting from the domain shift. 

These comparisons show that translations produced by all algorithms do improve upon the baseline with \uvc being the best performer in both translation directions. 
Notably, \uvc achieves a greater than $80\%$ reduction in $\ell_1$ error over the baseline on average for the \realtofake translation.
Comparing the result in \autoref{tab:metrics} shows that the translation quality measured in electron counts correlates strongly with those featuring ADC values. 
Additional evaluation of post-signal processing performance is provided in 
Appendix E.
%\ref*{supp:eval_full}.

For qualitative comparison, the signal-processed results for one sample \dst tile are shown in \autoref{fig:sigproc_ab}A for the \faketoreal translation and \autoref{fig:sigproc_ba}A for the \realtofake translation. 
The translated images in both figures are produced by the best performer, \uvc. 
Due to the mismatch in response functions, the signal-processed result for a domain \real sample may exhibit artifacts along the periphery of the tracks as exemplified by the area marked with the blue box. 
We anticipate an effective \faketoreal translation to replicate the artifact, while a \realtofake translation should eliminate it. 
This expectation aligns with the translations generated by \uvc, where the artifact is introduced in $\phi_{A}(\mathcal{G}_{A\rightarrow B}(a))$ and is removed in $\phi_{A}(\mathcal{G}_{B\rightarrow A}(b))$.

Considering the neural translation algorithm is trained in a purely data-driven fashion, i.e., solely based on the ADC waveform images without any input or constraints from physics or downstream applications, these are promising results.

\myhl{
\subsection{Domain shift mitigation for a supervised learning algorithm}
\label{subsec:eval_downstream}

\def\ecount{e}
\def\ecpred{E}

In this section, we investigate the effectiveness of UI2I translation techniques in mitigating the domain shift problem in a supervised learning context. Specifically, we design a supervised DL regression model to predict the number of ionized electrons from an ADC waveform. This model exhibits decreased performance when trained on domain $A$ and applied to domain $B$. Then, we test whether the UI2I translations can alleviate this degradation of the model performance.

The experiment proceeds as follows: we train predictive models---$\ecpred_A$, $\ecpred_B$, and $\ecpred_{\mathcal{G}}$---to estimate the total count of ionized electrons $\ecount$ in a waveform. Here, $\mathcal{G}$ is a neural translator that translates ADC waveform images from domain $A$ to domain $B$, such as \cyc, \acl, \uga, and \uvc. The models $\ecpred_A$, $\ecpred_B$, and $\ecpred_{\mathcal{G}}$ are trained using waveforms $a\in A$, waveforms $b\in B$, and translated waveforms $\mathcal{G}_{A\rightarrow{B}}(a)$ for $a\in A$, respectively.

After training, we evaluate all the models on waveforms from domain $B$. Due to the domain shift, we expect that $\ecpred_A$ will perform worse on $B$ than $\ecpred_B$. However, since $\ecpred_{\mathcal{G}}$ is trained on translated waveforms that closely resemble those in domain $B$, we expect that it will outperform $\ecpred_A$ when tested on $B$.
%All models utilize a network architecture similar to AlexNet~\cite{NIPS2012_c399862d}.
We use an AlexNet-like~\cite{NIPS2012_c399862d} architecture for the regression model. Further details on the model are provided in 
Appendix F.
%\ref*{supp:nn_for_electron_count_pred}.

\begin{table}[h!]
	\centering
	\caption{\textbf{Models trained on translated images mitigate the domain shift problem, as measured by MARE (lower is better).} $\ecpred_A$, trained on domain $A$ and applied to domain $B$, represents the worst-case scenario for domain shift. $\ecpred_B$, trained and tested on domain $B$, serves as the performance benchmark. The $\ecpred_\mathcal{G}$ models, trained on waveforms translated by a neural translator $\mathcal{G}$ and tested on domain $B$, demonstrate varying degrees of effectiveness in mitigating domain shift through UI2I translation methods.}
	\newcolumntype{P}{>{\raggedright\arraybackslash}p{.11\textwidth}}
	\begin{tabular}{l|PPPPPP}
		\toprule
		{} & $\ecpred_A$ & $\ecpred_{B}$ & $\ecpred_{\text{\cyc}}$ & $\ecpred_{\text{\acl}}$ & $\ecpred_{\text{\uga}}$ & $\ecpred_{\text{\uvc}}$ \\
		\midrule
		% $B$ &$0.389767$ & $0.210528$ & $0.221561$ & $0.223461$ & $0.257200$ & $0.216234$  \\
        $B$ &$0.390$ & $0.211$ & $0.222$ & $0.223$ & $0.257$ & $0.216$  \\
       \bottomrule
	\end{tabular}
	\label{tab:downstream}
\end{table}

\autoref{tab:downstream} summarizes the results of the evaluation of the regression models on domain $B$.
The performance is evaluated with mean absolute relative error (MARE), calculated as $n^{-1}\sum_{i=1}^{n}\left|(\hat{\ecount}_{i} - {\ecount}_{i})/{\ecount}_i \right|$, where $\hat{\ecount}_{i}$ represents the predicted total electron count and $n$ is the number of test examples.

As expected, $\ecpred_A$ exhibits the worst performance due to the domain shift. $\ecpred_B$ performs the best, as it was trained and tested on data from the same domain. The four $\ecpred_\mathcal{G}$ models, trained on translated waveforms, show varying degrees of effectiveness in mitigating the domain shift problem. Notably, $\ecpred_{\text{\uvc}}$ outperforms others, achieving comparable MARE to $\ecpred_B$. This aligns with our earlier findings from the pixel-wise difference analysis in \autoref{subsec:eval_metrics}. 
}

\section*{Discussion and future research direction}
Findings from this work highlight the potential of UI2I translation algorithms in addressing the challenges of domain shift in \tpc data. 
However, several issues require attention before these algorithms can be effectively used to translate between simulated and real detector data.

\paragraph{Scaling UI2I algorithms to work on large images.}
Existing UI2I translation algorithms have been developed and tested on images of size $(256, 256)$. 
Thus, the same-sized tiles are used in this study. However, full \tpc images of size $(800, 6000)$ are needed for downstream analyses. 
In applying the model to tiles and assembling them to form the full translated image, mismatches did occur along the tile boundaries. 
Therefore, as part of our future work, we need to develop network models and computational pipelines capable of handling full images.

\paragraph{Performing and Evaluating non-deterministic translations.} 
Another crucial aspect is the one-to-one nature of the translation. 
In this work, we addressed a problem where the translation between the two domains is fully deterministic and one-to-one.
However, in real detectors, multiple stochastic processes are present.
These stochastic processes will render the domain map non-deterministic, resulting in either one-to-many or even many-to-many relationships.
The non-determinism of the translation presents two challenges: 
1) how to adapt UI2I translation methods to handle non-deterministic mappings, and 
2) how to evaluate the quality of non-deterministic translations.

There are multiple ways to make a UI2I translation non-deterministic.
\acl~\cite{aclgan} presents one such approach, replacing a strong cycle-consistency constraint with a weaker adversarial consistency.
The DRIT family of models~\cite{lee2018diverse} demonstrates another method, separating the content (core part of the image that should be preserved) and the attributes (part of the image that changes during the translation). 
This separation allows for substituting multiple attributes for a single translation, resulting in a variety of output images.
BiCycleGAN~\cite{zhu2017toward} shows another interesting way to construct a one-to-many mapping  ($A\rightarrow B$) by adding an extra latent dimension $L$ to the \fake domain.
Then, it constructs a map $(A \times L \to B)$ which is a one-to-one map.
This method allows us to obtain a one-to-many translation ($A\rightarrow B$) by varying points in the latent dimension $L$.
The approaches presented by \acl, DRIT, and BiCycleGAN demonstrate that developing one-to-many and many-to-many translations is possible, indicating a promising direction for future research.

The shift to non-deterministic translations raises a question of how to evaluate the quality of the translation in a non-deterministic case.
In this work, we were able to construct a paired ground truth evaluation dataset due to the one-to-one nature of the problem.
This exact pairing allowed us to estimate the translation quality directly by comparing translated images to their ground truths.
However, in the case of one-to-many translation, constructing such a paired evaluation dataset becomes impossible.
Therefore, more sophisticated metrics are required to judge the quality of these non-deterministic translations.

We believe a robust evaluation protocol should focus on two aspects of the translation: 
1) realism, a neural translator's ability to replicate the distinctive features of the target domain during translation, and 
2) consistency, its capacity to translate without altering the underlying physical properties.
In non-scientific UI2I translation tasks, established metrics such as Fr\'{e}chet Inception Distance (FID)~\cite{heusel2017gans} and Kernel Inception Distance (KID)~\cite{binkowski2018demystifying} are commonly used to assess realism.
However, these metrics are based on the \texttt{InceptionV3} network~\cite{InceptionV3} pretrained on the \texttt{ImageNet} dataset~\cite{imagenet}, raising doubts about their applicability to scientific datasets.
On the other hand, translation consistency remains a relatively unexplored aspect of UI2I translation research. The exact definition of consistency in UI2I translation is likely dataset- and application-dependent. 
As far as we know, no established metrics or protocols exist for verifying such consistency, making this a critical area for future research in UI2I translation for scientific applications.

\paragraph{(Re)evaluation of systematic uncertainties in the presence of UI2I translation.} 
High-energy physics (HEP) experiments developed complex methods to estimate various systematic uncertainties affecting the final results (e.g.~\cite{abratenko2022search}).
Incorporating UI2I translation into the standard simulation chain may present unique challenges specific to HEP experiments.
These challenges are twofold: 1) estimating the systematic uncertainty (if any) stemming from the UI2I translation, and 2) understanding how the UI2I translation affects the already established systematic uncertainties.

While UI2I translation algorithms aim to bring the simulated (\fake) and experimental (\real) domains closer, they may introduce artifacts into the translation process.
These artifacts could be the source of additional systematic uncertainty, which may need to be quantified.
One possible approach to establishing the magnitude of such uncertainty is to train an ensemble of UI2I models and analyze the amount of variance introduced in the experimental results by the ensemble.
Alternatively, the UI2I translation could be treated as a ``detector calibration'' step, without assigning  specific systematic uncertainties to it. In this case, the uncertainties associated with UI2I would be incorporated into the uncertainties of other detector simulation parameters.

While UI2I translation techniques show promise in reducing the magnitude of systematic uncertainties, these reduced uncertainties still require careful evaluation. This evaluation process may involve a substantial effort to understand how UI2I translation algorithms interact with established methods of estimating systematic uncertainties in HEP experiments.
Further research is necessary to determine the most effective approaches for handling UI2I-related uncertainties in HEP experiments.

\section*{Conclusion}

In this work, we studied the potential of the UI2I translation algorithms to address the domain shift problem between simulation (domain \fake) and real data (domain \real) in the \tpc research.
We constructed a surrogate \tpc problem consisting of two simulated domains with a systematic difference in the detector response function.
This surrogate problem illustrates the typical source of the systematic uncertainty between the simulation and real data.
The deterministic nature of the detector response function allowed us to create a paired test dataset with the known ground truths for translations.

We tested four UI2I models (\cyc, \acl, \uga, \uvc) on the surrogate \tpc problem.
Our results show that the UI2I methods can successfully perform the translation of \tpc events as judged by pixel-wise metrics between the translation and the corresponding ground truth.
Notably, UI2I methods can identify and preserve the content of each event while translating its appearance. This indicates the feasibility of the application of the UI2I methods to translate \tpc data and improve the realism of the \tpc simulation.

Furthermore, we tested whether the obtained UI2I translations allow us to reduce the domain shift error of detector reconstruction algorithms, which are developed on simulation but applied to real data.
For this purpose, we employed a production-grade signal processing algorithm designed on simulation (domain \fake).
This algorithm experiences domain shift error when applied directly on domain \real.
However, we found that its domain error can be reduced by up to $80\%$ if we perform a UI2I translation ($B\rightarrow A$) before the application of the signal processing algorithm.
These results indicate that UI2I methods can be used for domain shift reduction in \tpc analysis.

Among the four tested UI2I models (\cyc, \acl, \uga, \uvc), the \uvc model achieves the best translation quality and introduces the fewest artifacts in the translated images.
This finding indicates that the \uvc model shows promise as a basis for more complex UI2I algorithms on scientific data.
To promote the reproducibility of our research, we publicly release the \dst dataset 
(\ifblinded\blinded\else\url{https://zenodo.org/record/7809108}\fi) 
and the code used in this study 
(\ifblinded\blinded\else\url{https://github.com/LS4GAN/uvcgan4slats}\fi).

While UI2I methods show promise in reducing the systematic differences between distinct domains of \tpc data and help to alleviate the domain shift error of the signal processing algorithm, there are several issues that remain to be addressed before their application becomes fully feasible.
First, the UI2I methods, currently developed on images up to 256 pixels in size, need to be scaled to work with larger images of up to 10,000 pixels.
Second, our work investigated a problem where the relationship between two domains is one-to-one. The actual relationship between the
\tpc detector simulation and real data is many-to-many. The performance of UI2I methods needs to be studied under many-to-many relationships.
Moreover, proper translation quality metrics need to be developed for the many-to-many case.
Finally, while UI2I methods may reduce the systematic difference between simulated and real data, one still needs to estimate potential systematic uncertainties introduced by these methods.
Likewise, work needs to be done to ascertain how the inclusion of the UI2I translation into the detector simulation pipeline may affect other systematic uncertainties.
Exploring these directions will be essential to fully leverage the potential of UI2I methods in \tpc research and broader scientific applications.

\section*{Acknowledgment}
\ifblinded
\blinded
\else
This work was supported by the Laboratory Directed Research and Development Program of Brookhaven National Laboratory, which is operated and managed for the U.S. Department of Energy Office of Science by Brookhaven Science Associates under contract No. DE-SC0012704.
\fi

\newpage
\bibliographystyle{unsrt}
\bibliography{bib.bib}

\appendix

\section{More details concerning the Simple Liquid-Argon Track Samples (\dst) dataset}
\label{supp:data}
Liquid Argon Time Projection Chamber (\tpc) detectors enclose a volume of liquid argon. 
As illustrated in 
Figure 2
%\autoref*{fig:lartpc-sigform} 
in the main text, energetic charged particles traversing the volume will ionize electrons from nearby argon atoms. 
Once these electrons are freed, they are made to drift to the readout side of the detector due to an applied uniform electric field. 
A \tpc detector readout is composed of several Anode Plane Assemblies (APAs). 
Each APA contains three sensitive wire planes. 
Each wire plane consists of an array of uniformly spaced parallel wires oriented at a unique angle. 
Electrons drift past the first two wire planes and are collected on the last wire plane. 
In this process, they induce electric current~\cite{ramo1939} in all nearby wires.
These currents are amplified, and the induced current waveforms are digitized to produce the Analog-to-Digital Converter (ADC) waveform images. 

The two-dimensional (2D) ADC waveform image from each wire plane provides a unique tomographic view of the distribution of ionized electrons. 
The horizontal axis of this image is the waveform sample time dimension, while the vertical axis is the wire channel dimension.
The pixel value of this image is the ADC value after the per-wire median ADC value has been subtracted.

\subsection{Idealization and simulated responses for \dst}
\label{supp:data-simulation}

A fully realistic \emph{simulation} requires a long chain of models for the following components: 
1) the initial flux of neutrinos or other particles of interest; 
2) the interaction cross sections, nuclear transport, and final state particle tracking through the volume; 
3) the production and drift of ionization electrons; 
4) the induction in the sensitive wire electrodes; and 
5) the final effects of electronic amplification and digitization. 

The models up to the production of ionized electrons form the first stage of the simulation, which encodes our understanding of particle physics. 
The remaining models form the second stage of the simulation, which encodes our understanding of detector physics. 
In constructing the \dst dataset, the difference between the two domains is designed to be caused by applying different detector response functions in the second stage. 
However, with the full simulation model chain, the resulting topology of ionized electron tracks is intricate. This intricacy can complicate the interpretation of subtle variations resulting from different detector response functions. 

To avoid this complication and simplify the software processing chain required to produce the \dst dataset, the full particle physics model chain (first stage) is replaced with a simplified, ideal-track model. 
This ideal-track model begins with the production of straight-line tracks randomly distributed in space and direction throughout the detector volume. 
Each track is made to ionize electrons at a rate corresponding to a minimum-ionizing muon. 
The result mimics the activity of cosmic muons traversing the detector.

After the simplified first stage, the latter stage employs the full detector physics model as implemented by the Wire-Cell toolkit~\cite{Qian_2018,wirecell} with an additional simplification that electronics noise (otherwise inescapable) is omitted. Though artificial, this choice allows for a focus on systematic differences due to disparate detector response functions.

The Wire-Cell toolkit software provides current state-of-the-art \tpc detector simulation and signal processing and is used by most \tpc experiments and prototype detectors either under construction or in operation today. 
The simulation components apply the effects of electron diffusion and absorption while transporting the ionization electrons through a uniform drift field in the bulk of the detector volume~\cite{LI2016160,CENNINI1994230}. 
Near the wire planes, ionization electrons are drifted through a far more complex electric field governed by the locations, sizes, and applied voltages of the sense wires. 
This detailed drift field and the associated Ramo weight fields~\cite{ramo1939} are provided to the toolkit as input. Here, we use fields calculated by the GARFIELD~\cite{garfield} software package via a \emph{2D model}~\cite{microboonesigproc2018p1} of the detector electrode arrays. 

As illustrated in 
Figure 3
% \autoref*{fig:data_simulation} 
in the main text, the \dst dataset's two domains are made unique by the diverse nature of these fields (quasi-one dimensional (1D) versus 2D). 
This work defines samples in domain \real as being produced with the aforementioned full 2D response function. 
On the other hand, samples from domain \fake are produced with a related yet different response function. 
The response function is obtained by masking the 2D response so that all contributions from regions near neighboring wires are removed. 
Comparing the illustration of the quasi-1D response with the 2D one in
Figure 3B 
%\autoref*{fig:data_simulation}B 
shows the quasi-1D response still is 2D in the remaining narrow region near the central wire, which explains the term ``quasi'' in the name.

Finally, after the electric current response, an electronics response and digitization model (linear scaling and truncation to $12$-bit integer) are applied. 
The final output from the simulation is the ADC waveform images that serve as the input to neural translators after passing through a few preprocessing steps.

\subsection{Data generation and preprocessing}
\label{supp:data_preprocessing}

To generate the \dst dataset, the simulation runs produced $10010$ events, each with $10$ ideal line sources at the minimum-ionization energy equivalent for muons. 
Each event results in a 2D ADC waveform image for each wire plane.
This work focuses only on the U plane, the first one the electrons encounter during their drift. 
The simulation employs a model of the ProtoDUNE-SP~\cite{pdspres2020} detector, which has six APAs at the readout. 
Hence, across the entire detector, the simulation produced a total of $60060$ U-plane images.

The image from the U plane is $800$ pixels in height and $6000$ pixels in width. 
The image height spans the electronics readout channels and provides a transverse tomographic view at a given time. 
The width denotes these samples over time. 

From each full readout image of shape $(800, 6000)$, we take a center crop of shape $(768, 5888)$. 
The center crop shape is chosen so it can be divided into tiles of shape $(256, 256)$, which typically are used as input to a neural translator.

In the conventional practice of analyzing \tpc readout images, it is common to apply similar center crops for various reasons, such as removing activity from background interactions originating outside the detector or providing a size more optimal for fast-Fourier transforms. 
Nevertheless, future work will investigate how to avoid this loss of information at the edge of the readout image.

In some instances, the randomness of placing $10$ ideal particles across the entire detector leads to one or more of the six APAs containing no ionization electrons. 
The resulting ``empty'' center crops of readout images are neglected, leaving $56,253$ non-empty center crops ($93.7\%$). 
From these non-empty center crops, $55,253$ crops are reserved for training and $1,000$ for testing.

Similarly, the sparseness of activity leads to a majority of $256\times256$ tiles being fully or nearly empty. 
We choose a threshold of $200$ pixels around the first local minimum of the distribution for domain \fake. 
To keep the tiles paired, we drop a pair if either domain \fake tile or its domain \real counterpart falls below the set threshold. 
After filtering, we have $1,065,870$ tile pairs for training and $18,887$ for testing. 

Of note, although the training dataset is paired, the UI2I translation training procedures shuffle both domains of the training dataset independently. 
Shuffling breaks the pairing, making the UI2I translation algorithms unable to benefit from the fact that the original dataset was paired.

\section{\uvc pretraining and training}
\label{supp:model-uvcgan-training}

The \uvc model used for \dst is identical to the one described in~\cite{torbunov2022uvcgan} except for three minor modifications: 
1) reducing the number of input/output channels to 1, 
2) removing all the normalization layers in the convolution blocks, and 
3) removing the output sigmoid activation from the generators.

Training the \uvc model on the \dst dataset consists of two stages: self-supervised pretraining and translation training. 
Although it is common practice to start the translation training directly with randomly initialized generators, there is evidence showing that initializing the generators by pretraining them on a simpler task provides an advantage over random initialization~\cite{torbunov2022uvcgan}. 
This study uses an image inpainting task to pretrain the generators. 
First, each \dst tile is subdivided into a grid of patches of size $(32, 32)$. 
Then, each patch is randomly masked by zeros with a probability of $.4$. 
The generators are pretrained to recover the masked regions, allowing them to learn nontrivial dependencies between different parts of a \dst image. 

Here, both generators are pretrained for $16,384,000$ iterations on the image inpainting task, configured similarly to~\cite{torbunov2022uvcgan}. 
A smaller learning rate of $6.25 \times 10^{-6}$ is used because \dst data have a larger range compared to natural images. 
Nonetheless, generators pretrained with this method failed to recover the full width of the tracks. 
Instead, they fill masked regions with very narrow tracks. 
We speculate this happens because pixel values away from the track cores are quite small compared to those near the cores. 
Therefore, their proper reconstruction gives a small benefit in terms of the $\ell_2$ loss. 
On the other hand, before the network learns to reconstruct these small-valued pixels properly, it is going to make many mistakes, which are costly in terms of the $\ell_2$ loss. 
The high cost of the mistakes compared to the small benefit of proper reconstruction creates a potential barrier to learning the full width of the tracks.

To lessen that learning barrier, we modify the $\ell_2$ loss function and reduce the penalty for the network to overwrite zeros incorrectly by $\alpha$. 
More precisely, let $y$ be an image from either domain \fake or \real and $\hat{y}$ be the inpainting output. 
The reconstruction loss then is defined as follows:
\begin{equation}
    \label{eq:bkg_penalty}
    L_\text{reco}\left(\hat{y}, y\right) = \frac{\alpha\cdot\sum_{y_{i,j}=0}\hat{y}^2_{i,j} + \sum_{y_{i,j}\neq0}\left(\hat{y}_{i,j} - y_{i,j}\right)^2}{H \times W},
\end{equation}
where $H$ and $W$ are the image height and width. 
During pertaining, we keep $\alpha$ at $0$ for the first $819,200$ iterations, allowing the network to freely overwrite the empty space without penalty. 
Then, we linearly anneal $\alpha$ to $1$ during the subsequent $2,457,600$ iterations.  
When $\alpha = 1$, the loss function in Equation~(\ref{eq:bkg_penalty}) reduces to the normal $\ell_2$ and is kept that way until the end of pretraining. 
An ablation study shows the modified $\ell_2$ loss expedited learning of the reconstruction of small-valued pixels. 
The generators trained with the modified $\ell_2$ loss also achieve a $\sim10\%$ lower reconstruction error than generators trained with the normal $\ell_2$. 

The translation on the \dst dataset was trained for $200$ epochs with $5000$ randomly selected tiles per epoch ($10^6$ iterations in total). 
We note that using slightly unequal initial learning rates for the generators ($10^{-5}$) and the discriminators ($5 \times 10^{-5}$) improves performance. 
The learning rates are kept constant for the first $100$ epochs and linearly annealed to zero during the second $100$ epochs. 

We also perform a hyperparameter (HP) optimization on coefficients of cycle-consistency loss, $\lambda_a$ and $\lambda_b$, and the discriminator gradient penalty parameters, $\lambda_\text{GP}$ and $\gamma$. 
The evaluation results presented in the work have been produced using the best model found in the optimization with $\lambda_a = \lambda_b = 1$, $\lambda_\text{GP}=1$, and $\gamma = 10$. 
Identity loss also is used for translation training with coefficients kept at half of $\lambda_a$ and $\lambda_b$. 
A more detailed discussion about loss coefficients and gradient penalty can be found in~\cite{torbunov2022uvcgan}.

\section{Modification and hyperparameter tuning for other \cyc-like models}
\label{supp:model-other}

This work required model modification and HP tuning of three other \cyc-like UI2I translation algorithms: \cyc~\cite{zhu2017unpaired}, \acl~\cite{aclgan}, and \uga~\cite{ugatit}.
Because all three algorithms originally were designed for photographic image translation, they use \texttt{tanh} at the final layer to limit the pixel value within $[-1, 1]$.
To adapt the models for the integer-valued \dst data, the final \texttt{tanh} activations are removed.

For \textbf{\cyc}, we conduct a grid search on two key HP values: generator architecture and the coefficient for the cycle consistency loss. 
We evaluate the ResNet generator with nine blocks and the U-Net generator with size $256$ input. 
We chose three cycle-consistency loss coefficient levels: $1$, $5$, and $10$ (default). 
As \cyc trains both generators jointly, we train six models (in total), one for each generator type and cycle consistency level. 
For each model, we train on $5000$ images (with batch size $4$) for $200$ epochs, which means a total of one million images are used for training.

For \textbf{\acl}, we employ three HP settings, one for each of the three unpaired translation tasks (selfie-to-anime, male-to-female, and eye-glasses removal) studied in~\cite{aclgan}. 
Because \acl does not train translations in both directions jointly, we train a total of six models, one for each translation direction and HP setting.
Each model is trained with a batch size of $4$ for $250000$ iterations. 
Again, a total of one million images are used for training. 
\acl can generate a variable number of outputs, each with a randomly generated style.
To compare directly with other algorithms, we have generated only one output and used $1$ for the random seed.

For \textbf{\uga}, we tune the cycle-consistency loss coefficient ($\lambda_{\text{cyc}}$) at three levels: $1$, $5$, and $10$ (default). 
Following the \uga default, we retain the identity consistency loss coefficients equal to those for the cycle consistency. 
Because \uga also trains both translation directions jointly, we train three models (in total). 
Each model is trained with a batch size of $4$ for $250000$ iterations, so one million total images are used for training.

\section{More details regarding signal processing}
\label{supp:sigproc-detail}

As detailed in \ref{supp:data}, the \tpc detects particles by recording ionization electrons produced along the particles' trajectories. 
These electron counts serve as the basis for deriving various parameters of the original particles, including momentum and mass. It is important to note that the \tpc readout, represented as ADC waveforms, does not directly provide the electron counts. 
Instead, it captures the digitized electric current they induce on the APA wires. 

In practice, the bipolar nature of \tpc ADC waveforms obscures an accurate and precise measurement of the underlying distribution of ionization electrons. 
To reveal this distribution so physically meaningful parameters about the original particles can be reconstructed (e.g., their momentum and mass), a procedure generically called \textit{signal processing} is applied.

Briefly, signal processing has two stages: deconvolution and high-pass filtering.
First, it performs a deconvolution of an ADC readout image with a model of the same detector response used in the simulation but averaged over each region near a wire.
The bipolar nature of the response inevitably causes the deconvolution to amplify low-frequency noise. 
To counter that, the second stage applies an adaptive high-pass filter known as \textit{signal region-of-interest (ROI) selection}. 

Due to the inevitable amplification of noise, signal processing is designed to contend with realistic detector noise by applying various filters. 
The interplay of the input noise, filters, and thresholds to define ROI makes signal processing especially sensitive to the presence of noise or the lack thereof.
The absence of noise in the \dst dataset causes the signal processing algorithm to fail.
Thus, post-processing of the noise-free \dst ADC waveforms is performed to add a realistic noise component. 
To do this, we linearly scale ADC pixel values to be consistent with the voltage levels originally produced by the amplifiers in the electronics prior to digitization. 
We then add noise generated from a model that has been previously developed to match observations of \tpc detectors. 
Finally, we rescale (re-digitize) the result back to ADC levels, and the signal processing can then be correctly applied.

Please refer to~\cite{microboonesigproc2018p1,microboonesigproc2018p2} for a more in-depth understanding of the signal processing procedure.

\section{More evaluation of translation quality}
\label{supp:eval_full}

\begin{table}[ht]
	% ================== parameters ==================
	\setlength\tabcolsep{5pt}
	\addtolength{\extrarowheight}{\belowrulesep}
	\aboverulesep=0pt
	\belowrulesep=0pt
	% ================== parameters ==================
	
	\newcommand{\cc}{\cellcolor{lightgray!50}}
	\centering
	\caption{Translation quality on ADC waveforms is evaluated in terms of $\ell_1$ and $\ell_2$ errors.}
	\begin{tabular}{rl|rr|rr}
		\toprule
		{} & {} & \multicolumn{2}{r|}{\fake to \real} & \multicolumn{2}{r}{\real to \fake} \\
		\cmidrule{3-6}
		algorithm					& HP variant & $\ell_1$ & $\ell_2$ & $\ell_1$ & $\ell_2$ \\
		\midrule
		\arrayrulecolor{black!50}
		\multirow{6}{*}{\cyc}	& (ResNet, $1$)  & $0.266$ & $6.123$ & $0.202$ &  $5.180$ \\
								& (ResNet, $5$)  & $0.171$ & $2.947$ & $0.235$ &  $5.449$ \\
								& (ResNet, $10$) & $0.147$ & $2.469$ & $0.322$ & $10.451$ \\
								& (UNet, $1$)    & $0.089$ & $0.177$ & $0.056$ &  $0.114$ \\
								& (UNet, $5$)    & $0.078$ & $0.178$ & $0.062$ &  $0.147$ \\
								& \cc (UNet, $10$)   & \cc $0.074$ & \cc $0.180$ & \cc $0.061$ &  \cc $0.159$ \\
		\midrule
		\multirow{3}{*}{\acl} & anime HP & $0.219$ & $5.476$ & $0.180$ &  $5.188$ \\
								& gender HP & $0.079$ & $0.727$ & $0.065$ &  $0.330$ \\
								& \cc glasses HP & \cc $0.083$ & \cc $0.566$ & \cc $0.039$ & \cc $0.121$ \\
		\midrule
		\multirow{3}{*}{\uga} & $\lambda_{\text{cyc}} = 1$  & $0.086$ & $1.367$ & $0.069$ &  $0.997$ \\
								& \cc $\lambda_{\text{cyc}} = 5$  & \cc $0.078$ & \cc $1.187$ & \cc $0.073$ &  \cc $1.161$ \\
								& $ \lambda_{\text{cyc}} = 10$ & $0.079$ & $1.404$ & $0.075$ &  $1.217$ \\
		\midrule
		\multicolumn{2}{c|}{\textbf{\uvc}}    & $\mathbf{0.030}$ & $\mathbf{0.033}$ & $\mathbf{0.025}$ & $\mathbf{0.027}$ \\
		\arrayrulecolor{black}
		\bottomrule
	\end{tabular}
	\label{supp_tab:metrics}
\end{table}

\begin{table*}[t]
	\setlength\tabcolsep{5pt}
	\addtolength{\extrarowheight}{\belowrulesep}
	\aboverulesep=0pt
	\belowrulesep=0pt
	\centering
	\caption{Translation quality on electron counts obtained from applying a signal processing procedure is evaluated in terms of mean absolute percentage difference (\%).}
	\begin{tabular}{r|c|c}
		\toprule
		{}       & \fake to \real & \real to \fake\\
		\midrule
		baseline & \multicolumn{2}{c}{$1.904$} \\ 
		\midrule
		\cyc & $1.220$ & $0.735$ \\
		\acl & $1.014$ & $0.582$ \\ 
		\uga & $0.998$ & $0.713$ \\
		\textbf{\uvc} & $\mathbf{0.549}$ & $\mathbf{0.391}$ \\
		\bottomrule
	\end{tabular}
	\label{tab:mampd}
\end{table*}

Here, we provide two additional evaluations of the translation quality. 
First, \autoref{supp_tab:metrics} depicts the $\ell_1$ and $\ell_2$ errors on ADC values for all HP settings discussed in \ref{supp:model-other}. 
The best performers for each algorithm are highlighted. 

Second, we evaluate signal processing results with pixel-wise percentage difference (PD). 
PD is especially useful for post-signal processing translation quality evaluation because the electron count distribution has a much broader range than ADC values along with a long heavy tail. Mathematically, for two scalars $x, \bar{x} \geq 0$, 
\[
    \text{PD}(x, \bar{x}) = \left\{
        \begin{array}{ll}
        \frac{\bar{x} - x}{\paren{\bar{x} + x}/2} \times 100\%& \text{if }\bar{x} + x > 0, \\
        0 & \textrm{if }\bar{x} + x = 0,
        \end{array}
        \right.
\]
and the mean absolute PD between two tensors is defined as the average of the absolute value of entry-wise PDs. 

\autoref{tab:mampd} shows the mean absolute PD averaged over $1000$ randomly selected test examples from \dst. 
Denote the signal processing procedure as $\phi$ and let $a\in$\fake and $b\in$\real be $a$'s counterpart. 
The baseline is defined as the mean absolute PD between $\phi(a)$ and $\phi(b)$. 
For a UI2I translation algorithm, we calculate the mean absolute PD between $\phi\paren{\mathcal{G}_{A\to B}(a)})$ and $\phi(b)$ for \faketoreal translation and between $\phi\paren{\mathcal{G}_{B\to A}(b)})$ and $\phi(a)$ for the \realtofake translation. 
\autoref{tab:mampd} indicates all neural translators offer an improvement over the baseline with those translations produced by \uvc achieving the best performance. 

\myhl{

\def\ecount{e}
\def\ecpred{E}

\section{Training details of the electron-count estimator $\ecpred$}
\label{supp:nn_for_electron_count_pred}
We designed the electron count predictor $\ecpred$ with the following architecture. The neural network consists of $5$ convolutional blocks followed by $2$ linear blocks. Each convolutional block includes a convolutional layer with a kernel size of $3$ and padding of $1$, followed by a leaky rectified linear unit (Leaky ReLU) activation function and an average pooling layer that halves the spatial dimensions (width and height). The first convolutional layer has $1$ input channel and $16$ output channels. In the subsequent convolutional layers, the number of output channels doubles with each block until it reaches $64$. The output of the convolutional blocks is then flattened before being passed through the linear blocks.

Each linear block starts with a dropout layer with a probability of $0.2$, followed by a linear layer and an activation function. The Leaky ReLU activation is used for the first linear block, while the identity activation is used for the final output. The first linear layer transforms the $8 \times 8 \times 64 = 4096$ input features into $128$ output features, and the second linear layer maps these $128$ features to a single output.

To ensure a fair comparison across different predictors, we initialized all models using a random number generator with seed $2024$. Each model was trained for $500$ epochs with a batch size of $4$, utilizing $80\%$ of the $1000$ samples from Section 4.2 for training and the remaining $20\%$ for testing. The learning rate was initially set to $0.0001$ and was reduced by a factor of $0.95$ every $10$ epochs. We used the mean absolute error ($\ell_1$) as the loss criterion and optimized the models using the AdamW optimizer with parameters $\beta_1 = 0.9$, $\beta_2 = 0.999$, and a weight decay of $0.01$.
}

\end{document}